\documentclass{article}
\usepackage{spconf,amsmath,graphicx}
\usepackage{booktabs}
\usepackage{tabularx}
\usepackage{multirow}
\usepackage{graphicx}
\usepackage{svg}
\usepackage{caption}
\usepackage{float}
\usepackage{xcolor}
\usepackage{color,soul}

\title{Discrete Audio Representation as an Alternative to Mel-Spectrograms for Speaker and Speech Recognition}
\name{Krishna C. Puvvada, Nithin Rao  Koluguri, Kunal Dhawan, Jagadeesh Balam, Boris Ginsburg} %\thanks{abc}
\address{NVIDIA, USA}
\begin{document}
\ninept

\maketitle
\begin{abstract}
%Compression based audio tokenization has seen renewed interest driven by its potential to facilitate the application of text language modeling approaches in audio domain.
Discrete audio representation, \emph{aka} audio tokenization, has seen renewed interest driven by its potential to facilitate the application of text language modeling approaches in audio domain. To this end, various compression and representation-learning based tokenization schemes have been proposed. 
However, there is limited investigation into the performance of compression-based audio tokens compared to well-established mel-spectrogram features across various speaker and speech related tasks.  In this paper, we evaluate compression based audio tokens on three tasks: Speaker Verification, Diarization and (Multi-lingual) Speech Recognition.
Our findings indicate that 
(i) the models trained on audio tokens perform competitively, on average within $1\%$ of mel-spectrogram features for all the tasks considered, and do not surpass them yet. 
(ii) these models exhibit robustness for out-of-domain narrowband data, particularly in speaker tasks.
(iii) audio tokens allow for compression to 20x compared to mel-spectrogram features with minimal loss of performance in speech and speaker related tasks, which is crucial for low bit-rate applications, and 
(iv) the examined Residual Vector Quantization (RVQ) based audio tokenizer exhibits a low-pass frequency response characteristic, offering a plausible explanation for the observed results, and providing insight for future tokenizer designs. % Proposed model and training recipes will be open sourced 

\end{abstract}
\section{Introduction}
\label{sec:intro}

Autoregressive modeling of token sequences has proven to be a powerful approach in natural language processing (NLP)\cite{brown2020language_gpt3, touvron2023llama}. Inspired by the success of large language models in NLP, the field of speech and audio is undergoing a transformative phase. 
Transformer-based models are increasingly gaining prominence\cite{radford2022whisper}, driven by the interest for a foundational audio model capable of both speech comprehension and generation. Such models are being designed to address a wide array of speech and audio-related tasks, including but not limited to Speaker Recogntion and Diarization, Automatic Speech Recognition (ASR), Text-to-Speech (TTS), Speech-to-Text translation, Speech-to-Speech translation, and Auditory Question Answering etc.
However, a noteworthy difference between text and audio domain is the discrete nature of text tokens vs. continuous nature of audio. Audio discretization (tokenization) is one way to circumvent this distinction.  Current tokenization schemes for audio can be broadly categorized into two types: (a) Neural compression based tokens and (b) Semantic tokens. As the name suggests, the former are designed for audio compression where as the later are derived through clustering intermediate layer representations when training for speech representation learning \cite{hsu2021hubert, chung2021w2v, chen2022wavlm, zhang2023google_usm}. One class of foundational audio model efforts like VALL-E~\cite{wang2023valle}, VALL-E X~\cite{zhang2023valle-x}, VIOLA~\cite{wang2023viola}, MusicLM~\cite{agostinelli2023musiclm} use compression based tokens where as models such as AudioLM~\cite{borsos2023audiolm}, AudioPaLM~\cite{rubenstein2023audiopalm} employ semantic tokens. While semantic tokens are proven to be efficient for ASR as well as most speaker related tasks, the same cannot be said for compression based codes and thus is an open question.

Soundstream~\cite{zeghidour2021soundstream}, EnCodec~\cite{defossez2022encodec} and DAC~\cite{kumar2023dac} are some of the recently proposed neural compression models for audio tokenization. They are encoder-decoder based architectures relying on Residual Vector Quantization (RVQ) \cite{zeghidour2021soundstream} to discretize  audio. RVQ, a hierarchical quantization method, employs a series of codebooks, with each codebook approximating the residual signal from the preceding codebook. Finally, the target signal is reconstructed through the summation of quantizer outputs. In contrast to semantic tokens, compression tokens are trained with the primary objective of reconstructing the original audio in a task-agnostic fashion. Thus, they are ideally suited for tokenized representation of audio. Nonetheless, they are compression based tokens and might lose information (quantization noise). Therefore it is advisable to investigate the limits of their performance compared to well established mel-spectrogram features, which is the topic of this paper.
% \vspace{-0.001mm}

With EnCodec\cite{defossez2022encodec} as the candidate for compression based token representation for speech, we evaluate Speaker Verification, Diarization tasks and ASR tasks. We make the following contributions:
% \vspace{-0.3cm}
\begin{itemize}
\itemsep-0.05cm
    \item Systematically study compression based audio tokens and understand their limitations for a variety of speaker and speech tasks for the first time (to the best of our knowledge).
    \item Demonstrate that models trained using EnCodec tokens achieve competitive results, on average within 1\% of mel-spectrogram features, across all the tasks considered.
    \item Show that models built on EnCodec tokens exhibit robustness for out-of-domain narrowband data in speaker related tasks.
    \item Illustrate that audio tokens enable audio compression up to 20x with minimal loss of performance for speaker and speech related tasks.
    \item Reveal that RVQ based EnCodec tokenizer exhibits a low-pass frequency response offering insights for future audio tokenizer design.

\end{itemize}

\noindent Overall, our findings suggest that compression-based audio codes, while falling slightly short of mel-spectrogram features, present a viable discrete alternative for audio representation and modeling.

\section{Audio Representation}
\label{sec:experiments}
\vspace{-1pt}
% \subsection{Audio Representation}
% \label{ssub:audio_representation}
As most contemporary neural compression models \cite{zeghidour2021soundstream, defossez2022encodec, kumar2023dac} utilize RVQ quantizer as the basic building block, without loss of generality, we employ EnCodec~\cite{defossez2022encodec} as a representative candidate for compression based audio tokenizer. EnCodec is comprised of 32 codebooks, each containing 1024 code vectors with a size of 128. It operates on audio input sampled at 24kHz, converting it into audio tokens at a rate of 75Hz. Consequently, EnCodec transforms an audio segment with a duration of \emph{d} seconds into tokens of size $32$$\times$$(d*75)$. Tokens from distinct codebooks but at same time-stamp are aggregated post embedding layer in all our experiments. 

For baseline, we extract 80-dim mel-spectrogram (Mel-Spec) features using a 512 FFT, a Hann window, 25 ms window length, and 10 ms shift. Mel-spectrogram features are normalized across time per frequency channel before being used as input for the model.

We compare EnCodec tokens (also referred to as audio tokens subsequently, albeit loosely, in the context of this paper) and mel-spectrogram features for Speaker verification, Diarization, monolingual English (En) ASR and multi-lingual English-Spanish (En-Es) ASR tasks.  We employ state-of-the-art models, namely TitaNet~\cite{koluguri2022titanet} for verification, diarization, and Fast Conformer~\cite{fastconformer} for mono and multi-lingual ASR. 

TitaNet is a speaker embedding extractor model utilizing a ContextNet-based encoder and an attentive pooling decoder. It generates a fixed-length speaker embedding of 192 dimensions from variable-length speech segments. We employ TitaNet-L as described in \cite{koluguri2022titanet}, featuring 25.3 million trainable parameters and an additive angular margin (AAM) loss \cite{deng2019_aam}.
Fast Conformer\cite{fastconformer} is an efficient variant of the Conformer model\cite{gulati2020conformer}. In this study, we utilize the Fast Conformer Transducer Large model (116M) with 4x subsampling. Both monolingual (En) and multilingual (En-Es) models employ Sentencepiece Byte Pair Encoding (BPE)~\cite{kudo2018sentencepiece} with a vocabulary size of 1024.

\vspace{-0.4cm}
\section{Datasets}
\label{sec:datasets}
\vspace{-0.1cm}
\subsection{Training Data}
\label{ssec:training_data}
We use following datasets to train Speaker Embedding extractor (TitaNet) and ASR models (En Fast Conformer, En-Es Fast Conformer):
\begin{itemize}
\vspace{-0.1cm}
\itemsep-0.05cm
    \item TitaNet: Voxceleb1\&2~\cite{Chung2018} and NIST-SRE 2004-2008 (LDC2009E100)
    \item En Fast Conformer : 960 hrs of LibriSpeech(LS) ~\cite{panayotov2015librispeech}
    \item En-Es Fast Conformer: LS and 776 hrs of MLS~\cite{pratap2020mls} Spanish data.
\end{itemize}

\vspace{-0.4cm}
% \subsubsection{Evaluation Data}
\subsection{Evaluation Data - Speaker Verification \& Diarization}
\vspace{-0.2cm}
\label{ssec:eval_data}
We utilize the VoxCeleb1-O cleaned test trial file\footnote{https://www.robots.ox.ac.uk/~vgg/data/voxceleb/meta/veri\_test2.txt} and NIST-SRE 2019 eval trial files for speaker verification assessment. For speaker diarization evaluation, Diarization Error Rate (DER) is reported on three datasets: (a) AMI Corpus \cite{carletta2005ami}, comprising Lapel and MixHeadset audio subsets from the partition set \cite{bredin2020pyannote}. (b) NIST-SRE-2000 \cite{martin2001nist}, including all sessions from LDC2001S97. (c) CH109 \cite{canavan1997callhome}, utilizing a subset of CALLHOME American English speech (CHAES) with only two speakers. No additional development set is used for tuning clustering parameters or speaker embeddings in speaker verification and diarization.

\vspace{-0.4cm}
\subsection{Evaluation Data - Automatic Speech Recognition}
\vspace{-0.2cm}
We evaluate the English ASR model using the standard `clean' and `other' variants of the dev and test partitions from LibriSpeech. To assess the En-Es model, we include evaluations on the En LibriSpeech dataset and report further results on CallHome Test~\cite{canavan1997callhome}. For the Spanish model, we report results on the test sets of MLS, MCV7~\cite{ardila2019common}, VoxPopuli~\cite{wang2021voxpopuli}, and Fisher~\cite{graff2010fisher}.

\section{Experiment Setup}
\label{sec:training_setup}
\subsection{Speaker Embeddings}
\vspace{-0.1cm}
We follow standard data pre-processing steps as mentioned in \cite{koluguri2022titanet}. For training speaker embedding extractor, longer speech segments ($>3$ seconds) are divided into random 1.5, 2, and 3-second chunks. The encoder input size is $T$$\times$$80$, where $T$ is the number of frames in the speech sample. 
%In all experiments where the input is audio codes, the embedding layer is initialized from EnCodec model and kept frozen during training.  

For each experiment to learn speaker embeddings we trained TitaNet for 100k steps on 4 nodes with 8 V100 32GB GPUs using a batch size of 64 on each GPU.  
Models are trained with a warm-up phase of 25K steps followed by cosine annealing and AdamW optimizer with a peak learning rate of 1e-3 and weight decay of 2e-3. For AAM loss we used margin($m$) as 30 and scale($s$) to 0.2. For a fair comparison of input features we haven't used any additional augmentation except Specaugment \cite{park2019specaugment} and trained for same steps without performing any optimal parameter selection. 
Evaluation of speaker verification systems employs EER and MinDCF with $P_{target}$ = $10^{-2}$ and $C_{FA}$ = $C_{Miss}$ = $1$. Both EER and DER use cosine similarity (CS) as the back-end similarity measure.

In diarization experiments, it's common to fine-tune clustering parameters and window settings using a development set. To maintain fairness, we did not employ additional data for parameter tuning. As reported in \cite{koluguri2022titanet}, we used different window sizes and shifts for speaker embedding extraction: 1.5 sec and 0.75 sec for AMI files, and 3 sec and 1.75 sec for CH109 and NIST SRE-2000 sets. A collar of 0.25 sec was used, and overlapping speech regions were excluded when calculating speaker error rates.

\subsection{Automatic Speech Recognition}
%hl{Krishna, Kunal edit this section accordingly based on above subsection.}
\vspace{-0.1cm}
For all ASR experiments, Fast Conformer transducer models are trained for 100k updates on 2 nodes with 8 A100 80GB GPUs each using a batch size of 64 on each GPU. We used AdamW with a peak learning rate of 2e-3 and weight decay value of 1e-3. We used 15K warmup steps with Cosine annealing and a minimum learning rate of 1e-6. Specaugment \cite{park2019specaugment} was employed in all experiments. 

In all experiments (both speaker and speech related) where the input is audio tokens, the embedding layer is initialized from EnCodec model and kept frozen during training. 

\begin{table}[ht]
\centering
\caption{Comparison of Equal Error Rate (EER \%) on popular speaker verification trial files, VoxCeleb1-Clean and NIST\_SRE 18. Models trained with EnCodec show robustness for out-of-domain narrowband data.}
\scalebox{0.8}{
    \centering
    \begin{tabular}{c|c|c|c}
         \toprule
         \textbf{Train} & \textbf{Feature} & \textbf{VoxCeleb1-O} & \textbf{\multirow{2}{*}{NIST-SRE 18}} \\
         \textbf{Dataset(s)} & \textbf{Type} & \textbf{Clean} & \\
         \midrule
         \midrule
         \multirow{2}{*}{VoxCeleb 1\&2} & Mel-Spec & 2.2 & 39.16 \\
         & EnCodec-32 & 2.90 & 27.72 \\
         \midrule
         VoxCeleb 1\&2 & Mel-Spec & 2.04 & 21.25 \\
         SRE & EnCodec-32 & 2.37 & 22.92 \\
         \bottomrule
    \end{tabular}
}
    \label{table:sv_main}
\end{table}

\vspace{-0.3cm}
\section{RESULTS}
\label{sec:results}
\vspace{-0.2cm}
\subsection{Speaker Verification}
\label{ssec:speaker_verification}
\vspace{-0.1cm}
We conducted two experiments to evaluate audio tokens for speaker verification. Initially, we trained two models: one using Mel spectrogram features and the other employing all 32 codebooks of EnCodec (EnCodec-32) . Both models were trained on VoxCeleb 1 \& 2 datasets sampled at 16 kHz and evaluated on VoxCeleb-clean (wide-band) and NIST-SRE (narrow-band) trial files. Results in Table \ref{table:sv_main} indicate that the EnCodec-trained model exhibits robustness and comparable performance to Mel-Spectrogram on VoxCeleb-clean trial file. It also demonstrates a significant 12\% absolute EER improvement on NIST SRE-18, despite not being originally trained for narrowband. To further investigate, we conducted an experiment where the TitaNet model was trained on additional narrowband data from NIST-SRE train set, revealing that the Mel spectrogram-trained model adapted better, displaying its adaptability to already seen domains. Meanwhile, the Encodec-trained model improved with additional data, maintaining similar performance in both narrowband and wideband scenarios.

\vspace{-0.2cm}
\subsection{Speaker Diarization}
\label{ssec:speaker_diar}
\vspace{-0.1cm}
Similar to speaker verification experiments we evaluate effect of input features on Speaker Diarization as well. Table \ref{table:sd_main} reveals that using speaker embeddings trained from wideband VoxCeleb 1\&2 and evaluating on both narrow and wideband diarization sets, Encodec demonstrates robust performance with an average DER of $5.08$, whereas models trained with Mel Spectrograms exhibit a DER of $6.00$. Similar to experiments in speaker verification, models trained with Mel Spectrograms outperform those trained with Encodec when utilizing both narrow and wideband datasets, highlighting their adaptability to similar domain data. Nevertheless, models trained with Encodec also show improvements, albeit not to the same extent as Mel spectrogram models.

\begin{figure}[t]
  \centering
  % \includesvg[width=0.82\columnwidth]{./figures/encodec_transfer_fn.svg}
  \includegraphics[width=0.82\columnwidth]{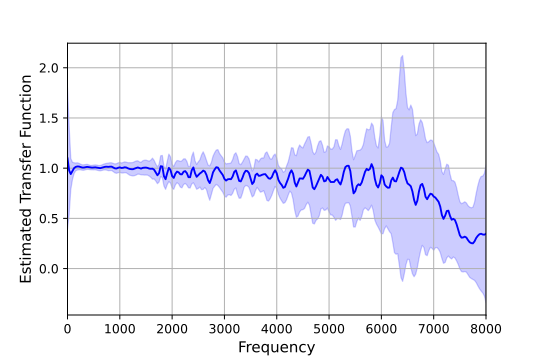}
  \caption{Estimated transfer function of EnCodec: Computed by comparing the spectral content of original audio and its compressed-decompressed version using EnCodec over 1000 random samples from VoxCeleb dataset}
  \label{fig:encodec_transfer_fn}
\end{figure}

To understand EnCodec token robustness on out-of-domain narrow-band data compared to Mel-Spectrogram features, exclusively trained on wide-band data, we conducted additional empirical analysis. We randomly sampled 1000 utterances from the VoxCeleb dataset and computed spectral content by marginalizing over the time dimension in two scenarios: original audio (FFT\_wav) and audio subjected to maximum bit-rate EnCodec compression and decompression using all codebooks (FFT\_EnCodec\_wav). We then derived the Estimated Transfer Function as \textit{FFT\_EnCodec\_wav / FFT\_wav}, as shown in Fig. \ref{fig:encodec_transfer_fn}. In an ideal EnCodec compression scenario, this function should consistently equal 1 across all frequencies, resulting in a flat line. However, we observed that the Estimated Transfer Function for EnCodec exhibits characteristics resembling a low-pass filter, attenuating frequencies beyond 6kHz. This phenomenon may elucidate the robustness of EnCodec tokens on out-of-domain narrow-band data, as even wide-band training data tokenized with EnCodec could appear narrow-band to the model due to this inherent low-pass nature.

\begin{table}[ht]
\caption{Diarization Error Rate (DER\%) on AMI-Lapel, AMI-MixHeadSet, NIST\_SRE 2000, and CH109 sets with oracle Speech Activity Detection (SAD) and estimated number of speakers. Last column shows average DER on all datasets.}
\scalebox{0.7}{
\begin{tabular}{c|c|c|c|c|c|c}
\toprule
\textbf{Train Dataset } & \textbf{Feature Type} & \textbf{\begin{tabular}[c]{@{}c@{}}AMI\\ Lapel\end{tabular}} & \textbf{\begin{tabular}[c]{@{}c@{}}AMI\\ MixHeadset\end{tabular} }& \textbf{\begin{tabular}[c]{@{}c@{}}NIST\_SRE\\ 2000\end{tabular}} & \textbf{CH109} & \textbf{AVG}  \\ 
\midrule
\midrule
\multirow{2}{*}{VoxCeleb 1\&2} & Mel-Spec & 3.11  & 2.41 & 12.70 & 5.81 & 6.00 \\
& EnCodec-32   & 4.12 & 3.82 & 10.19 & 2.19  & 5.08 \\ 
\midrule
VoxCeleb 1\& 2 & Mel-Spec & 3.07 & 2.38 & 8.40 & 1.37  & 3.80 \\
SRE & EnCodec-32 & 5.03 & 3.62 & 7.85 & 1.30  & 4.45 \\ 
\bottomrule
\end{tabular}
}
\label{table:sd_main}
\end{table}

\begin{table}[ht]
\caption{WER(\%) on Librispeech dev/test sets when trained on LS-960h with speech perturbation. EnCodec-32 shows same performance as Mel-Spec features on clean conditions but deteriorates slightly in 'other' conditions}
\centering
\scalebox{0.8}{
\begin{tabular}{c|c|c|c|c}
\toprule
\textbf{Feature} & \textbf{\multirow{2}{*}{dev-clean}} & \textbf{\multirow{2}{*}{dev-test}} & \textbf{\multirow{2}{*}{test-clean}} & \textbf{\multirow{2}{*}{test-other}} \\
\textbf{Type} & & & & \\
\midrule
\midrule
Mel-Spec& 2.12       & 4.88       & 2.27        & 5.03       \\ 

EnCodec-32 & 2.12       & 5.35       & 2.27        & 5.38       \\
\bottomrule
\end{tabular}
}
\label{table:asr_en_main}
\end{table}

\vspace{-0.5cm}
\subsection{Automatic Speech Recognition}
\label{ssec:asr}
\vspace{-0.1cm}
We evaluate EnCodec tokens for ASR in two different scenarios, monolingual (En only) and multi-lingual (En-Es). For monolingual scenario, we train two models, one using all 32 codebooks from EnCodec (EnCodec-32) and the other using mel-spectrogram features. Both models are trained on 3x speed perturbed (0.9x, 1x, 1.1x) LS-960h dataset and evaluated on clean/other portions of dev and test partitions of LS. Table \ref{table:asr_en_main} compares the (WER) performance of EnCodec-32 and Mel-Spec representations. We notice that EnCodec-32 performs on par with Mel-Spec features on dev-clean and test-clean, where as they exhibit slight deterioration on dev-other (4.88\% vs 5.32\%) and test-other(5.03\% vs 5.38\%). This shows that discrete representation of audio is a viable candidate for speech recognition. To further examine the EnCodec codes, we train multi-lingual En-Es models using EnCodec-32 and Mel-Spec features and evaluate on in-domain and out-of-domain test sets for both En and Es. From Table \ref{table:en-es_main}, we notice similar trends to monolingual experiments, i.e., while EnCodec-32 performs comparable to Mel-spec in clean conditions (ls-dev-clean and ls-test-clean in Table. \ref{table:en-es_main}), the performance falls short slightly for other conditions.

\begin{table*}[t]
\caption{WER(\%) for the En-Es model on various in- and out-of-domain (OOD) test sets. EnCodec exhibits similar performance as Mel-Spec on LS-clean sets but performs poorer on the LS-other sets. Mel-Spec outperforms EnCodec on the Spanish MLS sets. For the OOD eval, Mel-Spec comfortably beats EnCodec on En CallHome and Es MCV7, whereas Mel-Spec performs better on the hard Es Fisher set.}
\centering
\scalebox{0.8}{
    \begin{tabular}{c|c|c|c|c|c|c|c|c|c|c}
        \toprule
        & \multicolumn{5}{c|}{ \textbf{English} } & \multicolumn{5}{c}{ \textbf{Spanish} } \\
        \cline{2-11}
        \multirow{4}{*}{\textbf{Feature Type}}& \multicolumn{4}{c|}{\multirow{2}{*}{\textbf{In-Domain}} }  & \textbf{Out-of} & \multicolumn{2}{c|}{ \multirow{2}{*}{\textbf{In-Domain}}} & \multicolumn{3}{c}{\multirow{2}{*}{\textbf{Out-of-Domain}} } \\
        & \multicolumn{4}{c|}{}& \textbf{Domain} & \multicolumn{2}{c|}{}& \multicolumn{3}{c}{} \\
        \cline{2-11}
        & \textbf{LS-Dev} & \textbf{LS-Dev} & \textbf{LS-Test} & \textbf{LS-Test} & \textbf{CallHome} & \textbf{MLS} & \textbf{MLS} & \textbf{MCV7} & \textbf{VoxPopuli} & \textbf{Fisher} \\
        & \textbf{Clean} & \textbf{Test} & \textbf{Clean} & \textbf{Other} & \textbf{Test} &\textbf{ Dev} & \textbf{Test} & T\textbf{est} & \textbf{Test} &\textbf{ Test} \\
        \midrule
        \midrule
        Mel-Spec & 2.16&5.56&2.35&5.41&43.30&3.14&3.59&20.64&16.67&65.72 \\
        EnCodec & 2.16&5.95&2.36&5.89&45.32&3.60&4.02&25.08&16.97&63.67 \\
        \bottomrule

    \end{tabular}
}
\label{table:en-es_main}
\end{table*}

\vspace{-0.2cm}
\subsection{Ablations}
\vspace{-0.1cm}
To gain further insight into tokenized audio representation, we conduct two ablation studies. (i) Examine the performance of audio tokens at varying bit-rates and (ii) Compare EnCodec tokens with another recently proposed neural compression model (DAC) on En ASR task to examine potential differences originating from neural compression models themselves.  
\vspace{-0.3cm}
\subsubsection{Effect of bit-rate}
As RVQ based neural compressors typically employ a series of codebooks, the tokenized audio can be represented in varying bit-rates (depending on the number of codebooks used). Higher the bit-rate, higher the fidelity of representation. This offers a tradeoff between performance and bitrate (or compression ratio). We repeat Speaker verification and Diarization experiments by training TitaNet with varying number of codebooks (Tables \ref{table:ablation_sv_bitrate}, \ref{table:ablation_sd_bitrate}). Analogously, we repeat monolingual En ASR experiment with LS-960hrs with varying number of codebooks (Table \ref{table:ablation_asr_bitrate}).  We notice that audio tokens exhibit respectable performance even at 3 kbps (4 codebooks) with a 3.86\% EER on VoxCeleb1-Clean (Table \ref{table:ablation_sv_bitrate}) and an average DER of 5.4\% (Table \ref{table:ablation_sd_bitrate}). In En ASR experiment, audio tokens achieve 5.96\% WER using 6 kbps, i.e., 8 codebooks on test-other (Table \ref{table:ablation_asr_bitrate}). The performance deterioration on test-other is less than 1\% in terms of WER compared to mel-spec features at 6 kbps. In comparison, note that Mel-Spec features require at least 128 kbps (100 frames per sec $\times$ 80 features per frame $\times$ 16 bits per floating-point value). This suggests that audio codes can retain most of the performance even at low bit-rates and are good candidates for low bandwidth applications.

\begin{table}[t]

\centering
\caption{Comparison of Equal Error rate (EER \%) on VoxCeleb1-Clean and NIST SRE 18 with varying number of codebooks used to train speaker embedding model. Model trained with only 4 codebooks achieve acceptable performance on speaker verification task.}
\scalebox{0.8}{
\begin{tabular}{c|c|c|c}
    \toprule
    \textbf{Number of} &\textbf{Bit-rate} & \textbf{VoxCeleb1-O} & \textbf{\multirow{2}{*}{NIST-SRE 18}} \\ 
    \textbf{Codebooks} & \textbf{(kbps)} & \textbf{Clean} &  \\ 
    \midrule
    \midrule
    1      &0.75     & 10.71          & 33.59       \\
    2      &1.5     & 5.70           & 27.95       \\
    4      &3     & 3.86           & 25.38       \\
    8      &6     & 3.05           & 23.22       \\
    16     &12     & 2.61           & 23.62       \\
    32     &24     & 2.37           & 22.92       \\ 
    \bottomrule
    \end{tabular}
    }
\label{table:ablation_sv_bitrate}
\end{table}

% Please add the following required packages to your document preamble:
% \usepackage{graphicx}
\begin{table}[t]
\centering
\caption{Diarization Error Rate (DER\%) vs. number of codebooks  on AMI-Lapel, AMI-MixHeadSet, NIST\_SRE 2000, and CH109 sets with oracle Speech Activity Detection (SAD) and estimated number of speakers. Last Column shows the average DER (\%)}
\scalebox{0.78}{
\begin{tabular}{c|c|c|c|c|c|c}
\toprule
    \textbf{\begin{tabular}[c]{@{}c@{}}Number of\\ Codebooks\end{tabular}} & \textbf{\begin{tabular}[c]{@{}c@{}}Bit-rate\\ (kbps)\end{tabular}} & \textbf{\begin{tabular}[c]{@{}c@{}}AMI\\ Lapel\end{tabular}} & \textbf{\begin{tabular}[c]{@{}c@{}}AMI\\ MixHeadset\end{tabular}} & \textbf{\begin{tabular}[c]{@{}c@{}}NIST\_SRE\\ 2000\end{tabular}} & \textbf{CH109} & \textbf{AVG}   \\
    \midrule
    \midrule
    1 &0.75 & 16.15 & 18.9 & 14.11 & 4.39  & 13.39 \\
    2 &1.5& 8.56  & 10.85 & 8.81 & 2.03  & 7.56  \\
    4 &3& 6.57 & 5.39 & 8.18 & 1.44  & 5.40  \\
    8 &6& 6.30 & 6.06 & 7.87 & 1.39  & 5.40  \\
    16&12 & 4.28 & 4.11 & 7.86 & 1.39  & 4.41  \\
    32&24 & 5.03 & 3.62 & 7.85 & 1.3   & 4.45  \\ 
    \bottomrule
    \end{tabular}
}
    
\label{table:ablation_sd_bitrate}
\end{table}

\begin{table}[ht]
\caption{WER (\%) on Librispeech dev/test sets when trained on EnCodec tokens with varying bit-rates. Audio tokens exhibit respectable performance even at 6kbps bit-rate. }
\centering
\scalebox{0.8}{
\begin{tabular}{c|c|c|c|c|c}
\toprule
\textbf{Number of} & \textbf{Bit-rate} & \textbf{LS-Dev} & \textbf{LS-Dev} & \textbf{LS-Test} & \textbf{LS-Test} \\
\textbf{Codebooks} & \textbf{(Kbps)} & \textbf{Clean} & \textbf{Test} & \textbf{Clean} & \textbf{Other} \\
\midrule
\midrule
1                  & 0.75                                                                        & 8.61       & 24.34      & 8.22                      & 26.52                     \\
2                  & 1.5                                                                         & 3.42       & 10.62      & 3.42                      & 10.79                     \\
4                  & 3                                                                           & 2.44       & 7.13       & 2.60                      & 7.13                      \\
8                  & 6                                                                           & 2.23       & 6.02       & 2.35                      & 5.96                      \\
16                 & 12                                                                          & 2.26       & 5.77       & 2.45                      & 5.80                      \\
32                 & 24                                                                          & 2.16       & 5.68       & 2.30                      & 5.47                      \\ \hline
\end{tabular}
}

\label{table:ablation_asr_bitrate}
\end{table}

\vspace{-0.2cm}
\subsubsection{Audio Tokenizer}

\begin{figure}[t]
    \centering
    \includegraphics[width=0.85\columnwidth]{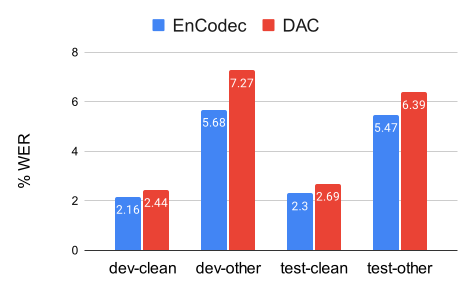}
    \caption{EnCodec vs DAC: WER (\%) comparison when trained on LS-960h dataset. EnCodec tokens show better over-all performance than DAC tokens.}
    \label{fig:encodec_dac_ls_asr}
\end{figure}

While we employ EnCodec as our tokenization candidate, the literature offers various other neural compression-based tokenization schemes. Here, we compare EnCodec with one such model, DAC~\cite{kumar2023dac}, which builds upon EnCodec and claims superior performance in speech reconstruction tasks. We train two models utilizing audio tokens from EnCodec and DAC for the En ASR task, using the LS-960 dataset. Fig.~\ref{fig:encodec_dac_ls_asr} shows that EnCodec outperforms DAC (5.47\% vs. 6.39\% WER on the test-other). This underscores the notion that not all audio tokens are equal, emphasizing the need for caution when applying them in downstream applications.

\section{CONCLUSION}
\label{sec:conclusion}

Using EnCodec as a candidate, we present the first systematic study of (RVQ) compression-based audio tokens across speaker verification, diarization and (multilingual) speech recognition tasks. Our observations indicate that, while audio tokens do not outperform mel-spectrogram features, they consistently demonstrate close competitive performance, often with in 1\%, thus establishing themselves as a viable tokenized alternative to continuous mel-spectrogram features. Furthermore, they exhibit robustness for narrowband out-of-domain data for speaker related tasks. They enable substantial audio compression with minimal loss of performance for all tasks considered. Finally, our study revealed an interesting low-pass nature of EnCodec, offering a plausible explanation as well as guidance for future codec design.  Code and models will be open sourced through NeMo~\cite{Harper_NeMo_a_toolkit}.

% \vfill
\bibliographystyle{IEEEbib}
\bibliography{strings,refs}

\begin{thebibliography}{10}

\bibitem{brown2020language_gpt3}
Tom Brown et~al.,
\newblock ``Language models are few-shot learners,''
\newblock {\em Advances in neural information processing systems}, vol. 33, pp. 1877--1901, 2020.

\bibitem{touvron2023llama}
Hugo Touvron et~al.,
\newblock ``Llama 2: Open foundation and fine-tuned chat models,'' 2023.

\bibitem{radford2022whisper}
Alec Radford, Jong~Wook Kim, et~al.,
\newblock ``Robust speech recognition via large-scale weak supervision,'' 2022.

\bibitem{hsu2021hubert}
Wei-Ning Hsu, Benjamin Bolte, et~al.,
\newblock ``Hubert: Self-supervised speech representation learning by masked prediction of hidden units,''
\newblock {\em IEEE/ACM Transactions on Audio, Speech, and Language Processing}, vol. 29, pp. 3451--3460, 2021.

\bibitem{chung2021w2v}
Yu-An Chung, Yu~Zhang, et~al.,
\newblock ``W2v-bert: Combining contrastive learning and masked language modeling for self-supervised speech pre-training,''
\newblock in {\em 2021 IEEE ASRU}. IEEE, 2021, pp. 244--250.

\bibitem{chen2022wavlm}
Sanyuan Chen, Chengyi Wang, et~al.,
\newblock ``Wavlm: Large-scale self-supervised pre-training for full stack speech processing,''
\newblock {\em IEEE Journal of Selected Topics in Signal Processing}, vol. 16, no. 6, pp. 1505--1518, 2022.

\bibitem{zhang2023google_usm}
Yu~Zhang, Wei Han, et~al.,
\newblock ``Google usm: Scaling automatic speech recognition beyond 100 languages,''
\newblock {\em arXiv preprint arXiv:2303.01037}, 2023.

\bibitem{wang2023valle}
Chengyi Wang, Sanyuan Chen, et~al.,
\newblock ``Neural codec language models are zero-shot text to speech synthesizers,''
\newblock {\em arXiv preprint arXiv:2301.02111}, 2023.

\bibitem{zhang2023valle-x}
Ziqiang Zhang et~al.,
\newblock ``Speak foreign languages with your own voice: Cross-lingual neural codec language modeling,''
\newblock {\em arXiv preprint arXiv:2303.03926}, 2023.

\bibitem{wang2023viola}
Tianrui Wang, Long Zhou, et~al.,
\newblock ``Viola: Unified codec language models for speech recognition, synthesis, and translation,''
\newblock {\em arXiv preprint arXiv:2305.16107}, 2023.

\bibitem{agostinelli2023musiclm}
Andrea Agostinelli, Timo~I Denk, et~al.,
\newblock ``Musiclm: Generating music from text,''
\newblock {\em arXiv preprint arXiv:2301.11325}, 2023.

\bibitem{borsos2023audiolm}
Zal{\'a}n Borsos, Rapha{\"e}l Marinier, et~al.,
\newblock ``Audiolm: a language modeling approach to audio generation,''
\newblock {\em IEEE/ACM Transactions on Audio, Speech, and Language Processing}, 2023.

\bibitem{rubenstein2023audiopalm}
Paul~K Rubenstein, Chulayuth Asawaroengchai, et~al.,
\newblock ``Audiopalm: A large language model that can speak and listen,''
\newblock {\em arXiv preprint arXiv:2306.12925}, 2023.

\bibitem{zeghidour2021soundstream}
Neil Zeghidour, Alejandro Luebs, et~al.,
\newblock ``Soundstream: An end-to-end neural audio codec,''
\newblock {\em IEEE/ACM Transactions on Audio, Speech, and Language Processing}, vol. 30, pp. 495--507, 2021.

\bibitem{defossez2022encodec}
Alexandre D{\'e}fossez, Jade Copet, et~al.,
\newblock ``High fidelity neural audio compression,''
\newblock {\em arXiv preprint arXiv:2210.13438}, 2022.

\bibitem{kumar2023dac}
Rithesh Kumar, Prem Seetharaman, et~al.,
\newblock ``High-fidelity audio compression with improved rvqgan,''
\newblock {\em arXiv preprint arXiv:2306.06546}, 2023.

\bibitem{koluguri2022titanet}
Nithin~Rao Koluguri, Taejin Park, and Boris Ginsburg,
\newblock ``Titanet: Neural model for speaker representation with 1d depth-wise separable convolutions and global context,''
\newblock in {\em ICASSP 2022}. IEEE, 2022, pp. 8102--8106.

\bibitem{fastconformer}
Dima Rekesh, Nithin~Rao Koluguri, et~al.,
\newblock ``Fast conformer with linearly scalable attention for efficient speech recognition,''
\newblock {\em arXiv:2305.05084}, 2023.

\bibitem{deng2019_aam}
Jiankang Deng, Jia Guo, et~al.,
\newblock ``Arcface: Additive angular margin loss for deep face recognition,''
\newblock in {\em CVPR 2019}, 2019, pp. 4685--4694.

\bibitem{gulati2020conformer}
Anmol Gulati, James Qin, et~al.,
\newblock ``Conformer: Convolution-augmented transformer for speech recognition,''
\newblock {\em Interspeech}, 2020.

\bibitem{kudo2018sentencepiece}
Taku Kudo and John Richardson,
\newblock ``Sentencepiece: A simple and language independent subword tokenizer and detokenizer for neural text processing,'' 2018.

\bibitem{Chung2018}
J.S. Chung, A.~Nagrani, and A.~Zisserman,
\newblock ``{VoxCeleb2}: Deep speaker recognition,''
\newblock in {\em Interspeech}, 2018.

\bibitem{panayotov2015librispeech}
Vassil Panayotov, Guoguo Chen, Daniel Povey, and Sanjeev Khudanpur,
\newblock ``Librispeech: an asr corpus based on public domain audio books,''
\newblock in {\em ICASSP 2015}. IEEE, 2015, pp. 5206--5210.

\bibitem{pratap2020mls}
Vineel Pratap, Qiantong Xu, et~al.,
\newblock ``Mls: A large-scale multilingual dataset for speech research,''
\newblock {\em arXiv preprint arXiv:2012.03411}, 2020.

\bibitem{carletta2005ami}
Jean Carletta, Simone Ashby, et~al.,
\newblock ``The ami meeting corpus: A pre-announcement,''
\newblock in {\em International workshop on machine learning for multimodal interaction}. Springer, 2005, pp. 28--39.

\bibitem{bredin2020pyannote}
Herv{\'e} Bredin, Ruiqing Yin, et~al.,
\newblock ``Pyannote. audio: neural building blocks for speaker diarization,''
\newblock in {\em ICASSP 2020}. IEEE, 2020, pp. 7124--7128.

\bibitem{martin2001nist}
Alvin Martin and Mark Przybocki,
\newblock ``The nist speaker recognition evaluations: 1996-2001,''
\newblock in {\em 2001: A Speaker Odyssey-The Speaker Recognition Workshop}, 2001.

\bibitem{canavan1997callhome}
Alexandra Canavan, David Graff, and George Zipperlen,
\newblock ``Callhome american english speech,''
\newblock {\em Linguistic Data Consortium}, 1997.

\bibitem{ardila2019common}
Rosana Ardila, Megan Branson, et~al.,
\newblock ``Common voice: A massively-multilingual speech corpus,''
\newblock {\em arXiv preprint arXiv:1912.06670}, 2019.

\bibitem{wang2021voxpopuli}
Changhan Wang, Morgane Riviere, et~al.,
\newblock ``Voxpopuli: A large-scale multilingual speech corpus for representation learning, semi-supervised learning and interpretation,''
\newblock {\em arXiv preprint arXiv:2101.00390}, 2021.

\bibitem{graff2010fisher}
David Graff, Shudong Huang, et~al.,
\newblock ``Fisher spanish speech (ldc2010s01),''
\newblock {\em Web Download. Philadelphia: Linguistic Data Consortium}, 2010.

\bibitem{park2019specaugment}
Daniel~S Park, William Chan, et~al.,
\newblock ``Specaugment: A simple data augmentation method for automatic speech recognition,''
\newblock {\em Interspeech}, 2019.

\bibitem{Harper_NeMo_a_toolkit}
Eric Harper, Somshubra Majumdar, et~al.,
\newblock ``{NeMo: a toolkit for Conversational AI and Large Language Models},'' .

\end{thebibliography}

\end{document}